\documentclass[prl,twocolumn,superscriptaddress,floatfix]{revtex4-1}
\newcommand{\ket}[1]{\ensuremath{\left|{#1}\right\rangle}}
\newcommand{\bra}[1]{\ensuremath{\left\langle{#1}\right|}}
\usepackage[sort&compress]{natbib}
\usepackage{graphicx, times}
\usepackage{color}
\usepackage{amssymb}
\usepackage{amsfonts}

\usepackage{subeqnarray}

\begin{document}

\title{Photon transfer in ultrastrongly coupled three-cavity arrays}

\author{S. Felicetti}
\affiliation{Departamento de Qu\'{\i}mica F\'{\i}sica, Universidad del Pa\'{\i}s Vasco UPV/EHU, Apartado 644, 48080 Bilbao, Spain}

\author{G. Romero}
\affiliation{Departamento de Qu\'{\i}mica F\'{\i}sica, Universidad del Pa\'{\i}s Vasco UPV/EHU, Apartado 644, 48080 Bilbao, Spain}

\author{D. Rossini}
\affiliation{NEST, Scuola Normale Superiore and Istituto di Nanoscienze - CNR, Pisa, Italy}

\author{R. Fazio}
\affiliation{NEST, Scuola Normale Superiore and Istituto di Nanoscienze - CNR, Pisa, Italy}
\affiliation{Centre for Quantum Technologies, National University of Singapore, Republic of Singapore}

\author{E. Solano}
\affiliation{Departamento de Qu\'{\i}mica F\'{\i}sica, Universidad del Pa\'{\i}s Vasco UPV/EHU, Apartado 644, 48080 Bilbao, Spain}
\affiliation{IKERBASQUE, Basque Foundation for Science, Alameda Urquijo 36, 48011 Bilbao, Spain}

\begin{abstract}
We study the photon transfer along a linear array of three coupled cavities where the central one contains an interacting two-level system in the strong and ultrastrong coupling regimes. We find that an inhomogeneously coupled array forbids a complete single-photon transfer between the external cavities when the central one performs a Jaynes-Cummings dynamics. This is not the case in the ultrastrong coupling regime, where the system exhibits singularities in the photon transfer time as a function of the cavity-qubit coupling strength. Our model can be implemented within the state-of-the-art circuit quantum electrodynamics technology and it represents a building block for studying photon state transfer through scalable cavity arrays.
\end{abstract}

\date{\today}

\maketitle

{\it Introduction}.---Light-matter interaction controls some of the most fundamental processes in nature~\cite{CohenBook}. Its study has allowed the development of impressive architectures where fundamentals of quantum mechanics can be tested~\cite{QuDevices}. 
The high level of control achieved on these setups~\cite{CavityQED,OpLattices,IonsReview}, motivated an increasing interest in the study of strongly correlated systems and collective phenomena. In particular, photon lattice models involving coupled cavity arrays have been devised in such a way that each cavity interacts with a single two-level system (qubit), thus realizing the Jaynes-Cummings-Hubbard model~\cite{Plenio2006, Hollenberg2006, Bose2007}. 
The possibility of implementing different geometries enables one to engineer quantum networks for distributed quantum information processing~\cite{Rempe2012}. These lattice models have proved useful to describe the scattering of a single-photon interacting with a qubit in a one-dimensional waveguide~\cite{Longo2010}. A similar approach can be used to consider homogeneously coupled cavity arrays~\cite{Nori2008}, where the central cavity-qubit interaction is in the strong coupling (SC) regime. 
However, when the rotating-wave approximation (RWA) ceases to be valid, the single-photon scattering properties have been only recently studied~\cite{Zhang2012}. Note that circuit QED technologies~\cite{Blais04, Chiorescu04, Wallraff04}, where photon lattice models have been proposed~\cite{Houck2012a, Hartmann2012,Houck2012b}, include disorder in the cavity coupling distribution due to imperfections. In addition, circuit QED has allowed the advent of the ultrastrong coupling (USC) regime of light-matter interaction~\cite{Bourassa09, Niemczyk10, Pol10}, described by the quantum Rabi model~\cite{Braak2011}. The latter displayed important consequences in strongly correlated systems, where a $\mathbb{Z}_2$ parity-breaking quantum phase transition was predicted~\cite{Tureci2012}.  

\begin{figure}[b]
\includegraphics[width=0.4\textwidth]{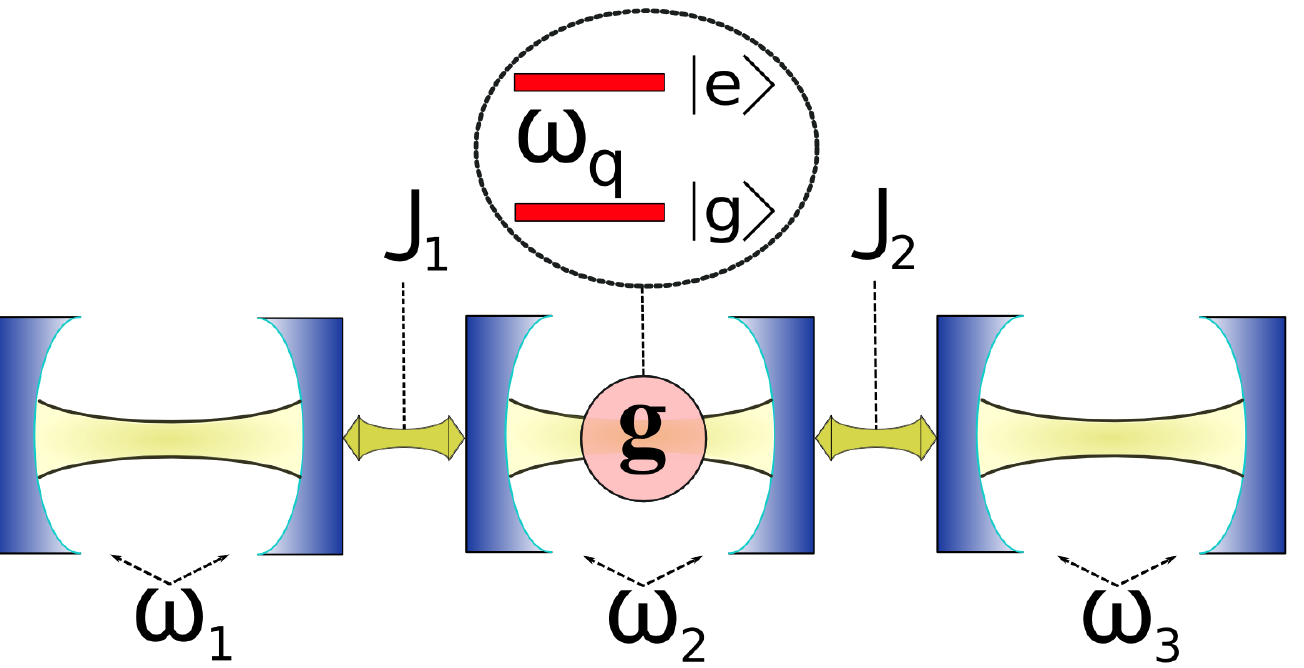}
\caption{\label{fig1} (Color online). Linear chain of three microwave cavities. The central cavity is coupled to a two-level quantum system in the strong coupling or ultrastrong coupling regime.}
\end{figure}

In this Letter, we consider the problem of photon transfer in a linear array of three coupled cavities, where a two-level system interacts at the central site in the SC and USC regimes (see Fig.~\ref{fig1}). 
This configuration can be thought as a microwave analogue~\cite{Saro2009} of the superconducting Josephson interferometer. We also include disorder in the cavity-cavity coupling, thus mimicking experimental imperfections. Under these conditions, we are able to unveil the following features: 
(i) in the SC regime, and for finite values of the hopping amplitudes and of the cavity-qubit coupling, a single excitation initially localized in the leftmost cavity is strictly forbidden to fully populate the rightmost cavity, similar to the delocalization-localization phenomenon~\cite{Blatter2010}; (ii) in the USC regime, the above restriction does not hold any more, and a single-excitation can fully populated the rightmost cavity for almost arbitrary Hamiltonian parameters; (iii) if the qubit transition energy is null, we find a complete and periodic excitation transfer from the leftmost to the rightmost cavity, independent of the cavity-qubit coupling strength. Our scheme represents a feasible building block~\cite{Houck2012a, Hartmann2012, Houck2012b, Cleland2011} to study photon excitation and state transfer towards scalable cavity arrays.

{\it The model}.---Our model consists of an array of three single-mode cavities, where the central site interacts with a two-level system in the SC or in the USC regime. A schematic representation is shown in Fig.~\ref{fig1} where each cavity is linked with its neighbor through a hopping interaction that, in general, is not weak enough to consider the RWA. The corresponding Hamiltonian reads 
\begin{equation}
  \begin{array}{rcl}
  H & = & \displaystyle \sum_{\ell=1}^3 \omega_{\ell} a^\dagger_{\ell} a_{\ell} + \frac{\omega_q}{2} \sigma_z 
  + g\sigma_x ( a_2^\dagger + a_2 ) \\ 
  & - & \displaystyle \sum_{\ell=1}^2 J_{\ell}(a^\dagger_{\ell} a _{{\ell}+1} + a_{\ell}^\dagger a^\dagger_{{\ell}+1} + {\rm H.c.} ),
  \label{fullHamiltonian}
  \end{array}
\end{equation}
where $a_{\ell} (a^{\dag}_{\ell})$ is the annihilation (creation) operator for photons on the $\ell$-th cavity (${\ell}=1,2,3$), $\omega_{\ell}$ being the characteristic frequencies and $J_\ell$ are nearest-neighbor hopping amplitudes. The qubit of frequency $\omega_q$ is located inside the central cavity and is described by the Pauli matrices $\sigma_\alpha$ ($\alpha = x,y,z$), while $g$ denotes the cavity-qubit coupling strength. The complexity of the Hamiltonian in Eq.~(\ref{fullHamiltonian}) is associated with the appearance of counter-rotating terms in the cavity-qubit and cavity-cavity interaction. 
Hereafter, we consider identical cavities ($\omega_{\ell}\!=\!\omega$), and the condition $J_{\ell}/\omega\!\sim\!0.1$ to assure a faithful numerical analysis of Hamiltonian~(\ref{fullHamiltonian}). We also set the energy scales in units of the resonator frequency $\omega$.

\begin{figure}[t]
\includegraphics[width=0.47\textwidth]{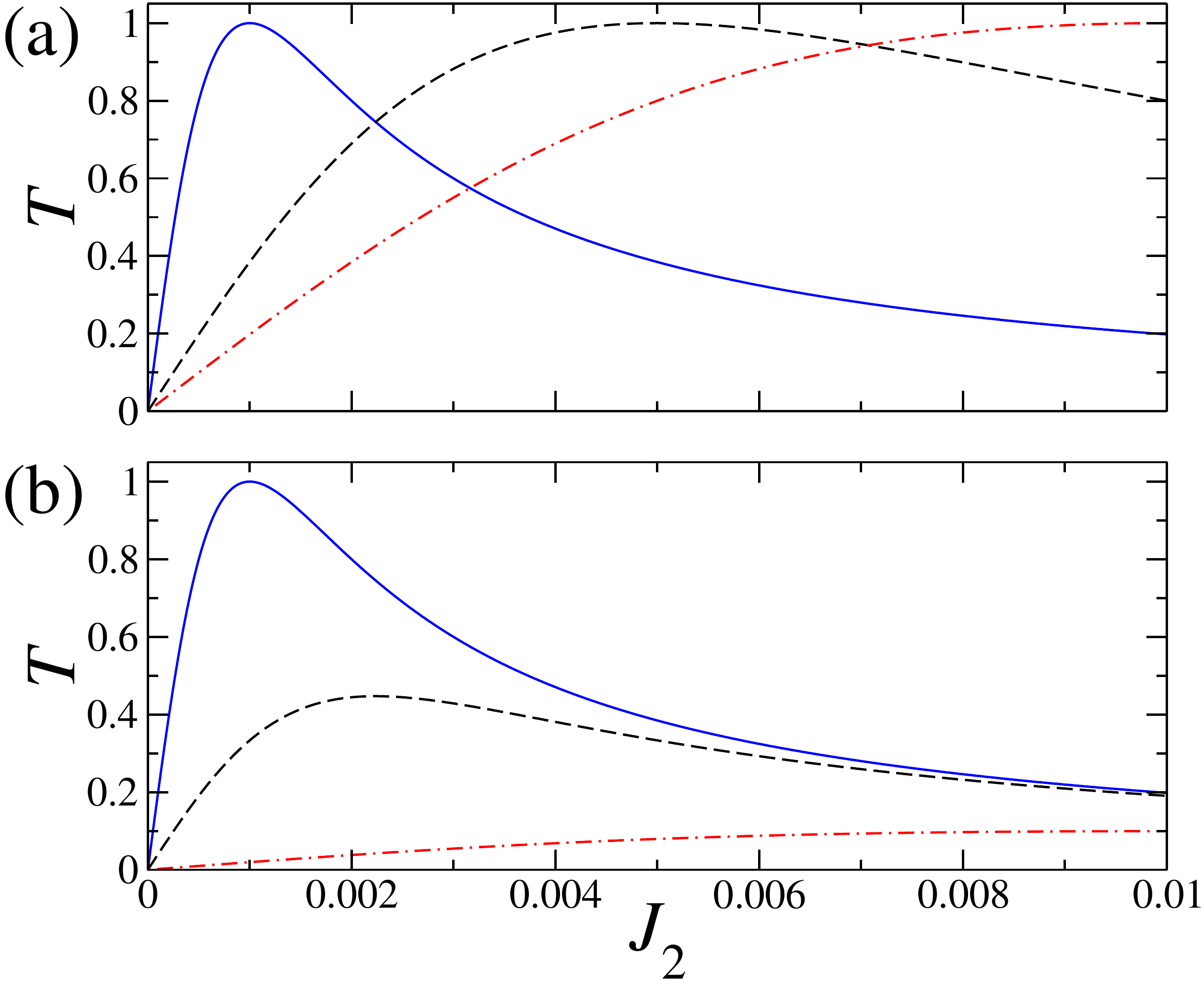}
\caption{\label{fig2} (Color online). Oscillation amplitude $T$ defined in Eq.~(\ref{amplitudeT}) as function of hopping parameter $J_2$ and for different parameter sets. In panel (a), for cavity-qubit coupling $g = 0$, we display $T$ for $J_1=0.001$ (continuous blue line), $J_1=0.005$ (dashed black line), and $J_1=0.01$ (dot-dashed red line). In panel (b), for  $J_1=0.001$, we consider $g = 0$ (continuous blue line), $g=0.002$ (dashed black line), and $g=0.01$ (dot-dashed red line).}
\end{figure}

The system is initialized in the state $\ket{\psi_0}\!\!=\!\!\ket{100} \otimes \ket{\rm g}$, corresponding to having a single photon in the leftmost cavity, zero in the others, and the qubit in its ground state. We address the dynamics dictated by the Hamiltonian~(\ref{fullHamiltonian}), and study the single-excitation transfer along our three-cavity array. Depending on the ratio $g/\omega$, two regimes can be identified: the SC regime for $g/\omega\!\lesssim\!0.1$, and the USC regime for $0.1\lesssim\!g/ \omega\lesssim\!1$.

{\it Single-photon transfer in SC regime}.---When $g/\omega\!\ll\!1$, the RWA provides a faithful description of the cavity-qubit dynamics, so that we can neglect the counter-rotating terms in Eq.~(\ref{fullHamiltonian}): $\sigma_x ( a_2^\dagger+a_2) \! \rightarrow \! ( \sigma_+ a_2 + \sigma_- a_2^\dagger)$. In the regime where hoppings $J_l/\omega\!\ll\!1$, the cavity-cavity RWA also holds, and the $U(1)$ symmetry provides the conservation of the total number of excitations. In this case, the time-evolution of the system will necessarily lead to a state of the form $\ket{\psi (t)}\!\!=\!\alpha \ket{000} \otimes \ket{\rm e} + \big( \beta \ket{100} + \gamma \ket{010} + \delta \ket{001} \big) \otimes \ket{\rm g}$. 
A full analytical solution can be found in the interaction picture and directly solving the Schr\"odinger equation.  At resonance, where qubit and resonator frequencies coincide ($\omega_q\!=\!\omega$), we can derive explicitly the probability amplitude for finding a photon in the rightmost cavity: $\delta(t)\!=T\left[ \cos{(\lambda t)} -1 \right]/2$, where $\lambda\!=\!\sqrt{g^2+J_1^2+J_2^2}$ and the amplitude reads
\begin{equation}
  T = \frac{2J_1 J_2 }{g^2+J_1^2+J_2^2}.
  \label{amplitudeT}
\end{equation}
The above result unveils a competition between cavity-cavity and cavity-qubit interaction. Figure~\ref{fig2}a indeed shows that, introducing disorder in the cavity-cavity couplings ($J_1\!\neq\!J_2$), the single-photon transfer has a counter-intuitive dependence on the hopping terms. In particular, for given parameters $g$, $\omega$, and $J_1$, the excitation transfer to the rightmost cavity exhibits a nonmonotonic behavior with increasing $J_2$ and is maximum for $J_1 = J_2$. Note that only in the homogeneous case ($J_1=J_2$) and for a negligible cavity-qubit coupling $g$, it is possible to have photon transfer with unit probability. On the other hand,  for any finite value of $g$, the latter is strictly forbidden. We also notice that, in Fig.~\ref{fig2}b and for fixed $J_1$, increasing $g$ decreases the amplitude $T$. This behavior has a simple explanation: when a single excitation tries to move from the left to the central cavity, it is scattered back by the cavity-qubit system without being fully absorbed. In fact, the probability for exciting the qubit $|\alpha|^2$ is inversely proportional to the square of the coupling strength.

{\it Single-photon transfer in the USC regime}.---If the coupling strength $g$ and the resonator frequency $\omega$ satisfy $0.1\lesssim\!g/ \omega\lesssim\!1$, the system enters the USC regime. In this case, photons are spontaneously generated from the vacuum such that the total number of excitations grows with the ratio $g/\omega$, enlarging unavoidably the associated Hilbert space~\cite{footnote0}. In order to provide a reliable system dynamics, we performed a fourth order-Trotter expansion of the evolution operator~\cite{SuzukiTrotter, footnote}. 

\begin{figure}[t]
\includegraphics[width=0.49\textwidth]{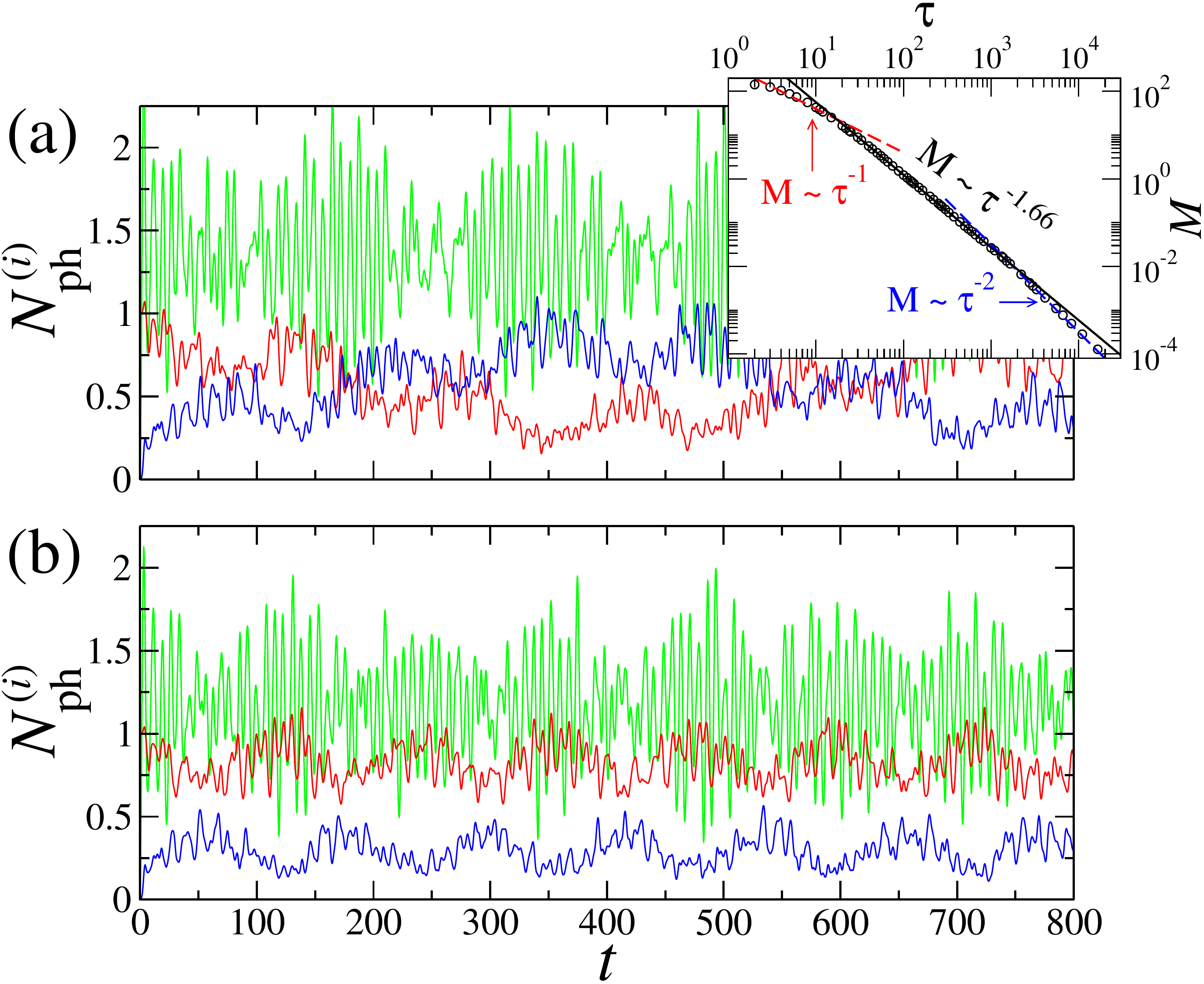}
\caption{\label{fig3} (Color online). Average photon number in each cavity  $N_{\rm ph}^{(i)} (t)$ at resonance condition ($\omega_q\!\!=\!\!\omega$), and for homogeneous cavity-cavity couplings ($J_1\!\!=\!\!J_2\!\!=\!\!0.1$). The red line stands for the leftmost cavity, the blue line for the rightmost  cavity, while the green line for the central cavity. The main panels refer to different cavity-qubit couplings: $g=0.9$ (a), $g=0.85$ (b). The inset shows the box counting analysis for $N_{\rm ph}^{(1)} (t)$ ($g = 0.9$), and displays $M$ as a function of $\tau$ (see the main text and Ref.~\cite{footnote2}). A power-law fit of the intermediate region gives a fractal dimension ${\cal D} \approx 1.66$.}
\end{figure}

Figure~\ref{fig3} shows the time-evolution of the average photon number in each cavity, $N_{\rm ph}^{(i)} (t) = \langle \psi(t) \vert a^\dagger_i a_i \vert \psi(t) \rangle$, starting from the state $\ket{\psi_0}$ that evolves according to Hamiltonian~(\ref{fullHamiltonian}). At first glance, one recognizes a highly irregular behavior of $N_{\rm ph}^{(i)} (t)$, which arises from the counter-rotating terms in the cavity-qubit interaction. Remarkably, this is developed by the unitary evolution of the system itself, and is not due to the limited time-resolution of our simulations. In order to quantify this behavior, we analyse the fractal dimension of $N_{\rm ph}^{(1)} (t)$ by using the modified box counting algorithm~\cite{BoxCount}. This consists in dividing the total time interval in segments of size $\tau$, and then covering the data with a set of rectangular boxes of size $\tau \times \Delta_i$ ($\Delta_i$ is the largest excursion of the curve in the $i$-th region $\tau$). One then computes the average excursion $M(\tau) = \sum_i \Delta_i/\tau$. The dimension ${\cal D}$ of the curve is defined by ${\cal D} = - \log_\tau M(\tau)$. One finds ${\cal D} = 1$ for a straight line, and ${\cal D} = 2$ for a periodic curve. Indeed, for times much larger than the period, a periodic curve uniformly covers a rectangular region. Any value of ${\cal D}$ between these integer values entails the fractality of the curve~\cite{footnote2}. In the case of Fig.~\ref{fig3}a, we obtained ${\cal D} \approx 1.66$ (inset to the figure). Furthermore, increasing $g$, we found a fractal dimension which rapidly decreases from ${\cal D} \sim 2$ (quasi-periodic curve) down to non integer values close to ${\cal D} \sim 1.5$ for large cavity-qubit couplings.

Two fundamental differences with respect to the SC regime can be found in the excitation transfer analysis. First, in the USC regime a single photon excitation can be completely transferred from the leftmost to the rigthmost cavity, also for a finite value of the coupling strength $g$ [see, {\it e.g.}, Fig.~\ref{fig3}a]. This is the typical situation for almost arbitrary Hamiltonian parameters in the USC regime, a disparity that can be explained as follows. In the SC regime, the initial state $\ket{\psi_0}$ has a finite overlap with the eigenstate $\ket{E} = (g\ket{100} \otimes \ket{\rm g} + J\ket{000} \otimes \ket{\rm e})/\sqrt{g^2+J^2}$,  which is conserved by the time-evolution. Consequently, the state $\ket{001} \otimes \ket{\rm g}$ is not accessible (as far as $g \neq 0$). On the other hand, in the USC regime the state $\ket{E}$ is no longer a system eigenstate. In fact, any eigenstate of Hamiltonian~(\ref{fullHamiltonian}) has a structure that covers the full system Hilbert space, thus enhancing the overlap with the state $\ket{001}\otimes\ket{\rm g}$. Consequently, complete transfer to the rightmost cavity is generally allowed.

\begin{figure}[t]
  \includegraphics[width=0.46\textwidth]{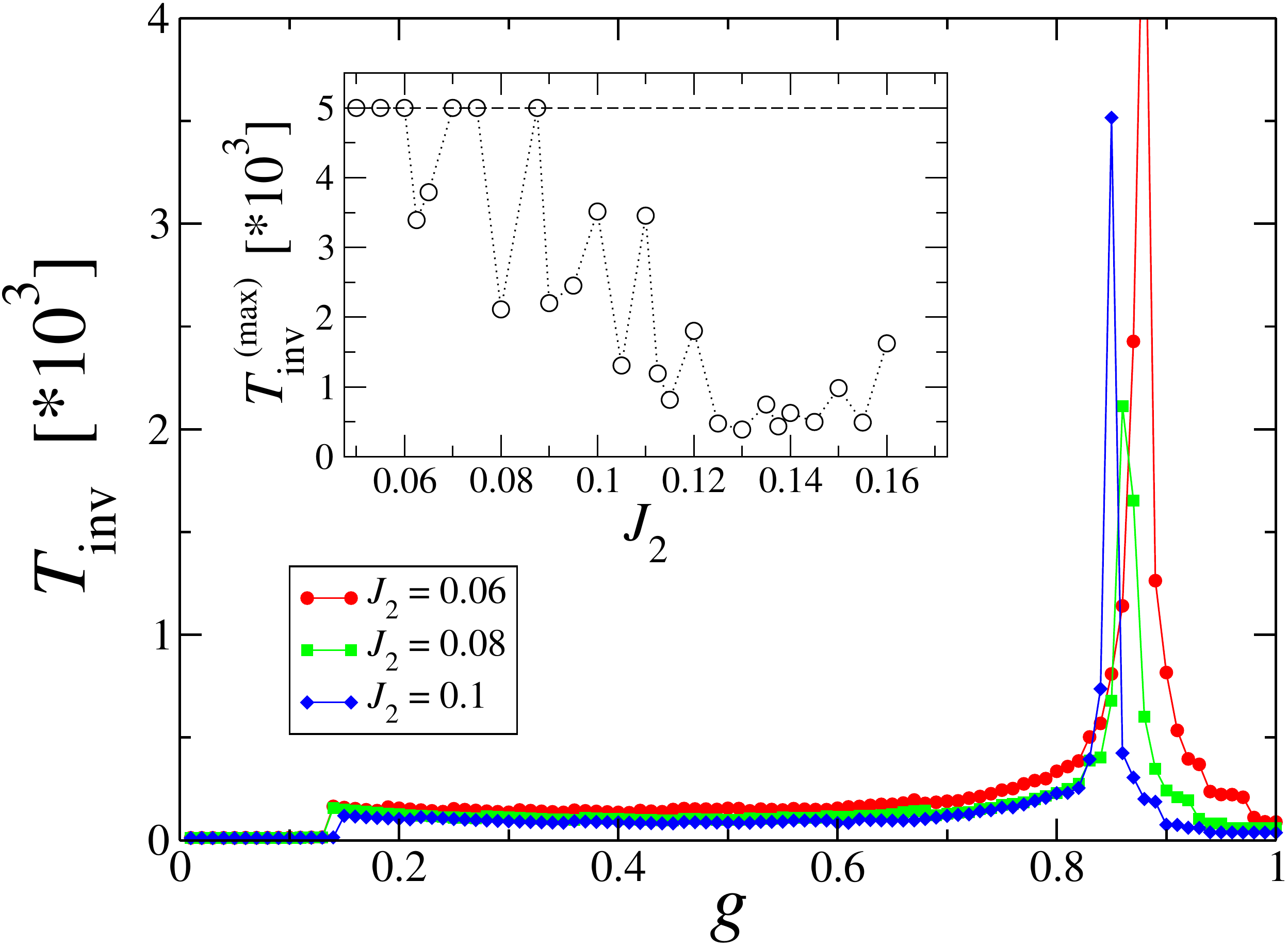}
  \caption{\label{fig4} (Color online). Population inversion time $T_{\rm inv}$ as a function of the cavity-qubit coupling constant, defined as the time in which $N_{\rm ph}^{(3)}(t)$ becomes bigger than $N_{\rm ph}^{(1)}(t)$ [that is, $\Delta N_{\rm ph} (T_{\rm inv}) = 0$ in Eq.~(\ref{deltan})]. Here, we have fixed $J_1 = 0.1$. Note the SC-USC transition for $g \approx 0.14$ and the inhibition of state transfer for $g = g_c \approx 0.85$. The inset displays the maximum value of the inversion time $T_{\rm inv}^{(\rm max)}$ that is reached at $g_c$, as a function of $J_2$~\cite{footnote}.}
\end{figure}

The second feature to be highlighted is that the photon transfer is strongly inhibited for a specific value of the cavity-qubit coupling strength, as displayed in Fig.~\ref{fig3}b. In particular, in Fig.~\ref{fig4} we analyzed the inversion time $T_{\rm inv}$, defined as the time where the average photon number becomes bigger in the rightmost cavity than in the leftmost cavity, as a function of the coupling strength $g$, and for different values of the cavity-cavity coupling strength. We found that there exists a critical value $g_c$, dependent on the system parameters, for which the time needed to observe single-photon transfer dramatically increases. For our set of parameters, fixing $J_1 = 0.1$ we found $g_c \approx 0.94 - 0.97 \times J_2$, while the inversion time $T_{\rm inv}^{(\rm max)}$ in correspondence to such value 
exhibits a quite irregular pattern of oscillations with $J_2$ (inset to Fig.~\ref{fig4}). To our knowledge, this is the first observation of this behavior in the excitation transfer properties of a cavity QED system. This phenomenon is specific to the USC dynamics and occurs in the {\it higher-coupling region} of the Rabi model ($g/\omega\!\gtrsim\!0.4$)~\cite{WolfVallone2012}, a zone where the photon production exceeds the RWA predictions, and an analytical treatment becomes difficult despite the integrability of the model~\cite{Braak2011}. 

We highlight the sudden increase of $T_{\rm inv}$ occurring in the {\it lower-coupling region} of the Rabi model, which is zoomed-in in Fig. 5. This abrupt behavior is due to the SC/USC regime crossover. If the RWA holds, the population inversion time is given by: $T_{\rm inv}=\arccos(1- \lambda^2/(J_1^2+J_2^2))/\lambda$, where  $\lambda$ has been defined  previously. For $g\!>\!\sqrt{J_1^2+J_2^2}$, the population inversion never occurs. Contrariwise, when the counter-rotating terms are taken into account, we observe population inversion at finite time also for larger value of the coupling $g$. For our three-cavity array, the parameter $g_t=g/\sqrt{J_1^2+J_2^2}$ provides us with an operational definition for the SC/USC transition of the cavity-qubit interaction. For small values of $g_t$, the RWA describes correctly the transfer dynamics. When $g_t>1$, the full model behavior differs quantitatively and qualitatively from the JC model predictions. We observe also that varying $J_2$ with respect to $J_1$ results in a longer population inversion time: in both regimes, inhomogeneity in the hopping terms hinders the excitation transfer.
\begin{figure}[t]
  \includegraphics[width=0.47\textwidth]{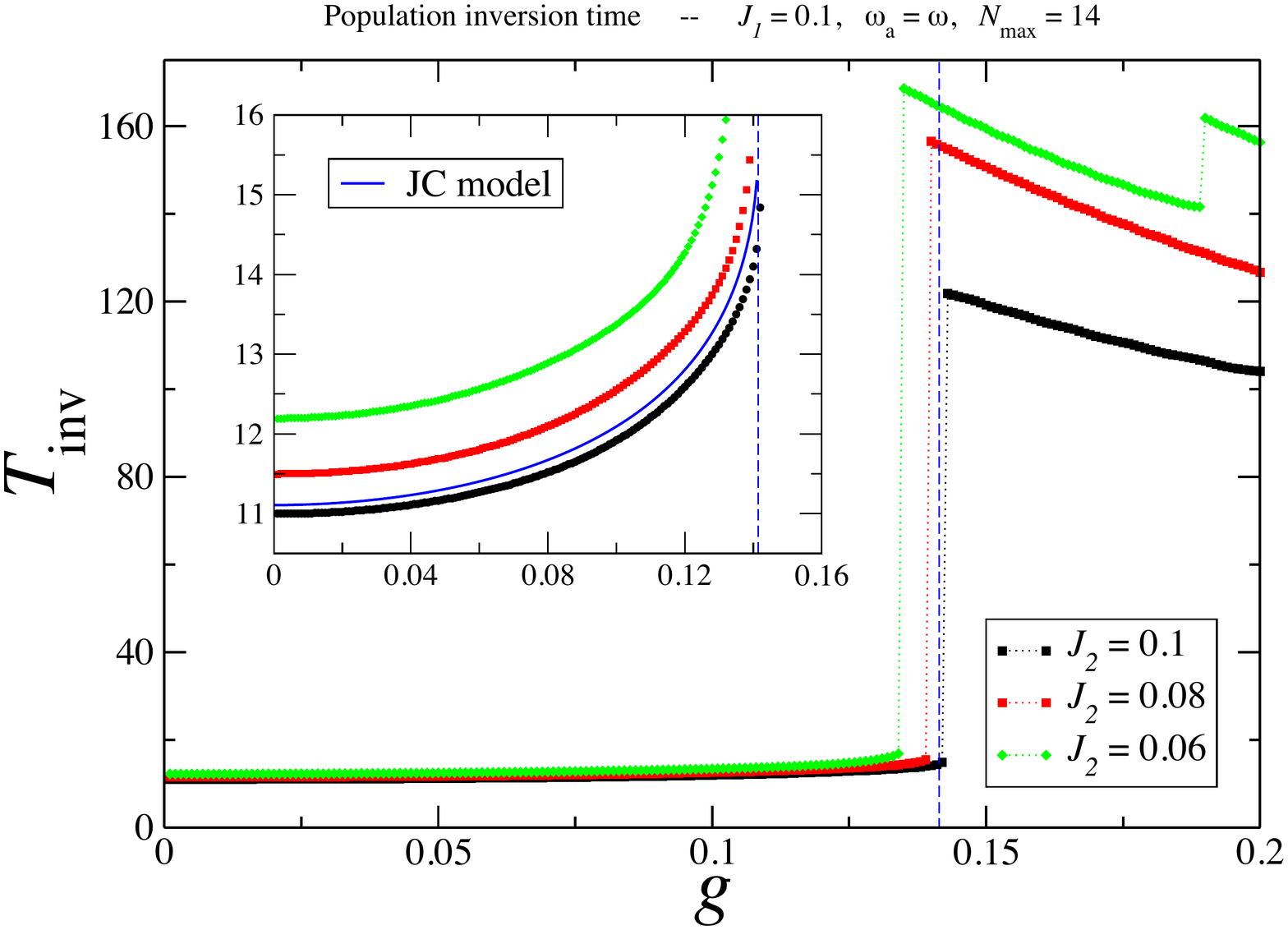}
  \caption{\label{fig5} (Color online). Population inversion time $T_{\rm inv}$ as a function of the cavity-qubit coupling constant $g$, and for fixed value $J_1=0.1$. The blue and the black lines correspond to the homogeneous case. The blue line corresponds to the analytical solution when the RWA holds for cavity-cavity and cavity-qubit interaction. The black, red and green lines are obtained through numerical simulations of the full model ruled by the Hamiltonian~(\ref{fullHamiltonian}).}
\end{figure}

We have also analyzed the transfer of coherent states, \ket{\phi_0}=\ket{\alpha}, and of arbitrary linear superposition states $\ket{\phi_{0}}=p\ket{0} + e^{i\theta}\sqrt{1-p^2}\ket{1}$, with $p$ randomly chosen in the interval [0,1]. We numerically simulated the system evolution, setting $\ket{\psi_{0}}=\ket{\phi_0}\ket{0}\ket{0}\ket{\rm g}$ as initial state, and we recorded the behavior of the transfer fidelity, defined as: $F={\rm Tr}[\rho_{0}\rho(t)]$, where $\rho_0=\ket{\phi_0}\langle\phi_0\vert$ and $\rho(t)$ is the state of the rightmost cavity at time $t$. The results we found are consistent with those relative to the case of single-excitation transfer. The interplay between the hopping constant $J$ and the cavity-qubit coupling strength $g$, rules the state transfer dynamics: with increasing $J$ the transfer is more likely to happen, while increasing $g$ results in smaller values of the transfer fidelity. Specifically, in the SC regime, as far as $g\neq0$, it is impossible to observe complete state transfer, i.e. $F\!<\!1$ at any time. This is not the case in the USC regime, where the transfer fidelity, be in a linear superposition or a coherent state, can be close to unity, also when $g$ and $J$ are of the same order. These results show the peculiar features of state transfer physics beyond the RWA, and may pave the way for developing a general theory in presence of USC regimes. Further details on state transfer features can be found in the supplemental material. 

{\it Degenerate qubit case}.---When the qubit frequency vanishes, a closed analytical solution of the dynamics governed by Hamiltonian~(\ref{fullHamiltonian}) is available if we consider the RWA in the cavity-cavity interaction. We may consider the quantum simulation of this model by means of a coupled cavity-qubit system in the SC regime, see Fig.~\ref{fig1}, and the application of a strong classical driving to the qubit~\cite{Solano2003,Ballester2012}. It can be shown that the dynamics will be ruled by the effective Hamiltonian
\begin{equation}
  {H_{\rm eff}} = \omega \sum_{\ell=1}^3 a^\dagger_{\ell} a_{\ell} + \frac{g}{2}\sigma_x ( a_2^\dagger+a_2 )
  - \sum_{{\ell}=1}^2 J_{\ell}(a^\dagger_{\ell} a _{{\ell}+1} + {\rm H.c.} ) \,.
\end{equation}
In this case, the excitation transfer exhibits a smooth periodic behavior in time, which is independent of the cavity-qubit coupling strength, and allows a complete transfer at regular times. In fact, the time evolution of the difference between the average photon number in the leftmost and rightmost cavities, starting from the initial state $\ket{\psi_0}\!\!=\!\!\ket{100} \otimes \ket{\rm g}$ in the homogeneous case, reads
\begin{equation}
  \Delta N_{\rm ph} (t) = N_{\rm ph}^{(1)}(t) - N_{\rm ph}^{(3)}(t) = \cos{(\sqrt{2}Jt)}.
  \label{deltan}
\end{equation}
This result holds both in the SC and in the USC regime, hence the counter-rotating terms do not modify the excitation transfer properties of the system. We point out that, when $g/\omega\!\gtrsim\!0.1$, spontaneous photon generation occurs: $\langle a_1^\dagger a_1\rangle$ and $\langle a_3^\dagger a_3\rangle$ keep a fractal time dependence, despite their difference follows a smooth and regular behavior.

{\it Conclusions and outlook}.---We studied excitation transfer in an array of three coupled cavities, where a two-level system interacts with the central cavity, focusing on the comparison between strong and ultrastrong coupling regimes of light-matter interaction. In the SC regime, the cavity-qubit interaction $g$ and the cavity-cavity inhomogeneities $J_1\!\neq\!J_2$ constrain the excitation transfer, thus inhibiting a complete population tunneling. On the contrary, in the USC regime a much richer scenario appears. A complete photon transfer is generally allowed, even for finite values of the cavity-qubit coupling strength and for inhomogeneous hoppings, although there exists a specific regime for which the tunneling rate becomes negligible. The complexity of the USC dynamics, that is generated by counter-rotating terms, manifests itself in the highly irregular time pattern of the observables, exhibiting a fractal behavior. Despite this fact, the physics beyond the RWA plays an important role for the enhancement of quantum state transfer. Finally, in the degenerate qubit case, the excitation transfer is regular and its period does not depend on the cavity-qubit coupling strength. 

The proposed scheme can be implemented with state-of-the-art current superconducting circuit technology or, alternatively, its quantum simulation may be considered. Furthermore, it can be used as building block for realizing large controlled networks for quantum simulation and distributed quantum information processing involving the quantum Rabi model in all coupling regimes.

{\it Acknowledgments}.---We acknowledge fruitful discussions with F. Ciccarello.
This work was supported by the Spanish MINECO FIS2012-36673-C03-02; UPV/EHU UFI 11/55; Basque Government IT472-10; SOLID, CCQED, PROMISCE, and SCALEQIT European projects.

\newpage 
\section{Supplementary material}

\begin{figure}[b]
\includegraphics[width=0.45\textwidth]{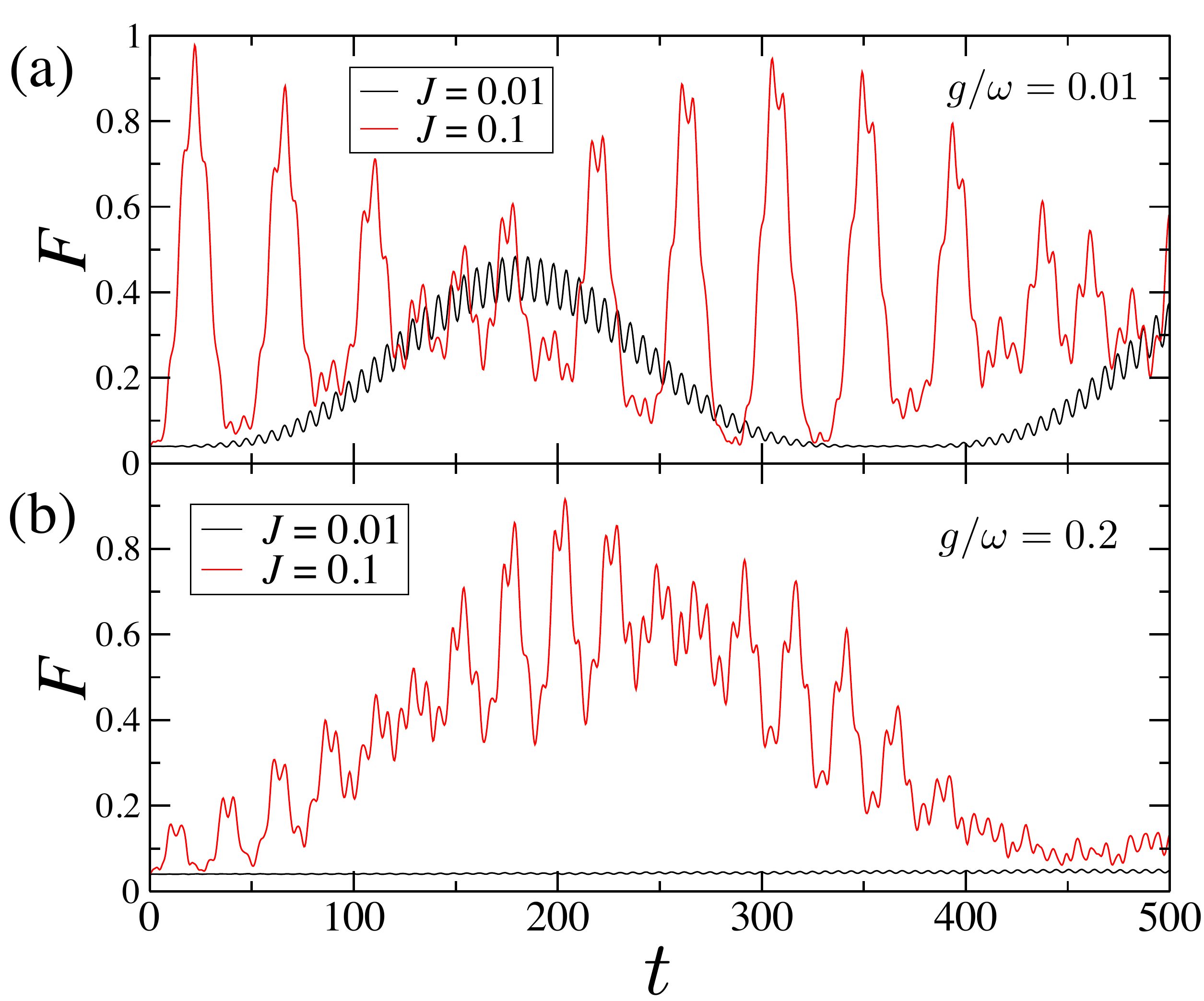}
\caption{(Color online) (a) State transfer fidelity $F={\rm  Tr}\left[ \rho_{0} \rho(t)\right]$ over time for different system parameters. We define $\ket{\phi_{0}}=p\ket{0} + e^{i\theta}\sqrt{1 - p^2} \ket{1}$, with $p=0.2$ and $\theta=0.63$. In the SC regime, that is when $g=J=0.01$ (black line), the state transfer fidelity is bounded: it cannot reach 1 as far as $g\neq0$. (b) When counter-rotating terms are involved in the dynamics ($J \gtrsim 0.1$ and/or $g\gtrsim 0.1$), the fidelity can be close to 1 even when $J<g$ (red line).}
\label{super}
\end{figure}

\subsection*{State transfer: linear superposition}
In this section, we consider the state transfer of a Fock state linear superposition $\ket{0}$ and $\ket{1}$, containing zero and one excitations of the cavity mode, respectively. For simplicity, we have considered homogeneous cavity-cavity couplings, $J_1\!\!=\!\!J_2$, and resonance condition, $\omega_q\!=\!\omega$. The system is initialized in state $\ket{\psi_{0}}=\ket{\phi_0}\ket{0}\ket{0}\ket{\rm g}$, where $\ket{\phi_0}\!=\!p\ket{0} + e^{i\theta}\sqrt{1-p^2}\ket{1}$, with $p$ randomly chosen in [0,1]. We define the state transfer fidelity as $F={\rm  Tr}\left[ \rho_{0} \rho(t)\right]$, which measures the overlap of the rightmost cavity state $\rho(t)$ with the leftmost cavity initial state $\rho_{0}=\ket{\phi_0}\bra{\phi_0}$.

\begin{figure}[t]
\includegraphics[width=0.5\textwidth]{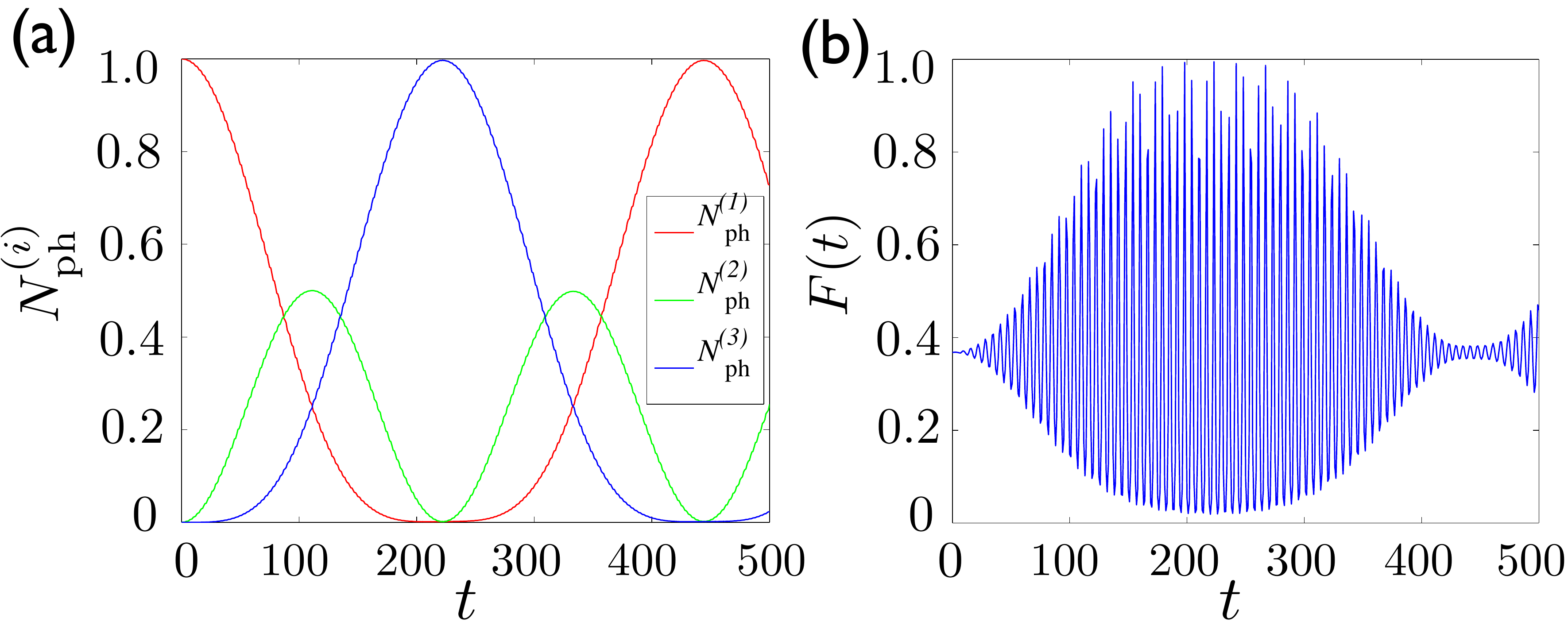}
\caption{(Color online) Transfer of a coherent state of amplitude $\alpha=1$ along a three-cavity array ($g=0$). (a) Average photon number for the leftmost (blue), central (green) and rightmost (red) cavities. (b) State transfer fidelity over time. In the case of qubit absence, the coherent state is fully transferred. Observe that the fidelity at time zero has a finite value $F\approx0.38$, which corresponds to the overlap between the vacuum state and the considered coherent state.}
\label{gzero}
\end{figure}

The results for different system parameters are shown in Fig.~\ref{super}. On one hand, in Fig.~\ref{super}a we have chosen a cavity-qubit coupling in the SC regime, $g/\omega=0.01$, and two values for the cavity-cavity coupling $J/\omega \! =\ 0.01,0.1$. Firstly, when both quantities are in the SC regime, the cavity-qubit interaction sets an upper bound for the maximum of the state transfer fidelity (black line). Secondly, if the cavity-cavity coupling is in the USC regime, $J/\omega=0.1$, then almost complete state transfer occurs (red line). On the other hand, in Fig.~\ref{super}b, we plot the fidelity $F$ for a cavity-qubit coupling in the USC regime, $g/\omega=0.2$, and $J/\omega \! = 0.01,0.1$ for the cavity-cavity coupling. When $g$ is one order of magnitude larger than $J$ (black line), the transfer is completely inhibited. Contrariwise, there is an enhancement in the state transfer when the cavity-cavity coupling is in the USC regime (red line).       

\begin{figure}[b]
\includegraphics[width=0.5\textwidth]{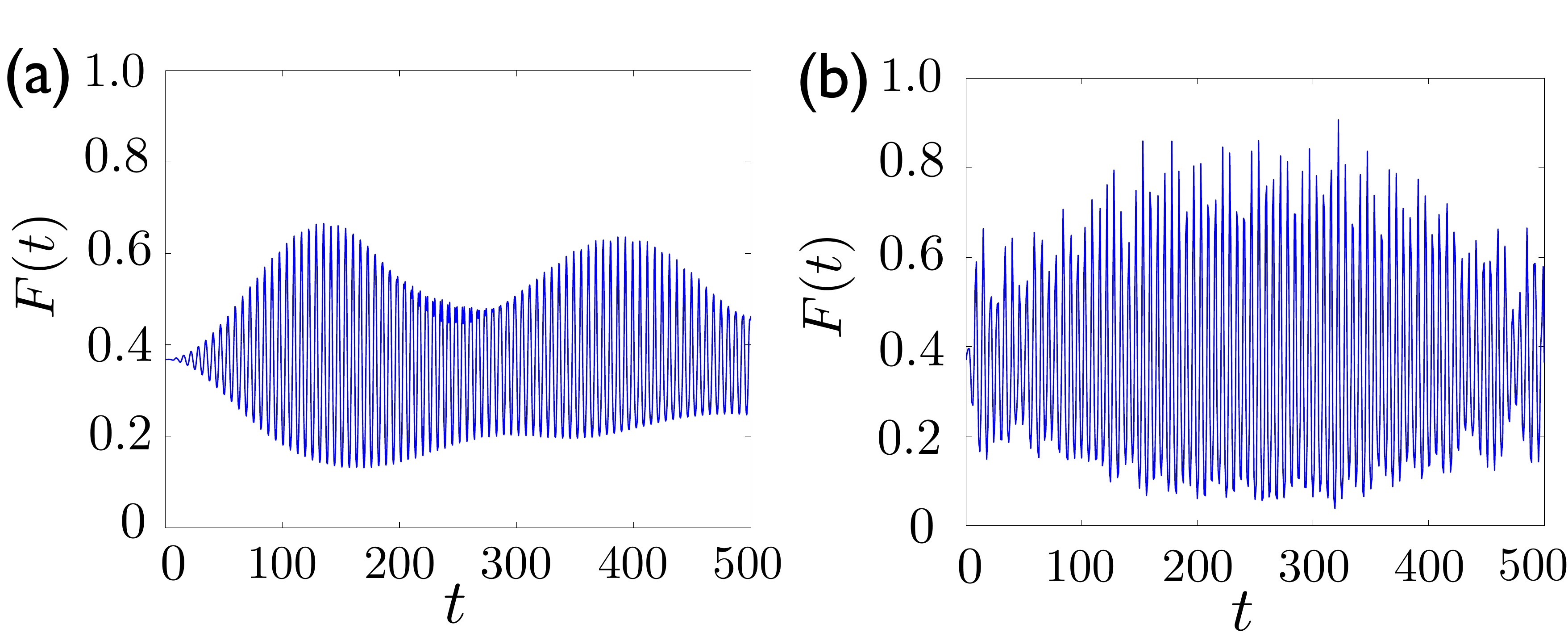}
\caption{(Color online) (a) State transfer fidelity over evolution time, for a coherent state of amplitude $\alpha=1$. The system parameters are given by $g=0.02$ and $J=0.01$ (SC regime). (b) State transfer fidelity over evolution time, for a coherent state of amplitude $\alpha=1$. The system parameters are given by $g=0.2$ and $J=0.1$ (USC regime). The maximum value that the fidelity can reach is not directly bounded by the cavity-qubit interaction.}
\label{coherSC}
\end{figure}

\subsection*{State transfer: coherent state}
In this section, we report the transfer properties of coherent states. We numerically simulated the system dynamics starting from the initial state $\ket{\psi_0}=\ket{\phi_0}\ket{0}\ket{0}\ket{\rm g}$, where $\ket{\phi_0}=\ket{\alpha}$ is a coherent state of amplitude $|\alpha|$. For the sake of clarity, we first show how a coherent state would transfer along a three-cavity arrays in absence of the qubit interaction. This corresponds to setting $g=0$ in our model. Such dynamics is shown in Fig.~\ref{gzero} for a coherent state with $\alpha=1$. In Fig.\ref{gzero}a, we show the average photon number for the three cavities, while Fig.~\ref{gzero}b contains the state transfer fidelity. These figures show that the coherent state crosses the central cavity and it is recomposed in the rightmost one. The fast-oscillating behavior of the fidelity is due to relative phase rotation of different coherent state components. We observe also that the fidelity oscillates around a value slightly smaller than $F=0.4$ due to the finite overlap between the coherent state $\ket{\alpha}$ and the vacuum state $\ket{0}$.

Let us now consider our full model, composed of three cavities and a qubit interacting with the central one.
The transfer of coherent states follows the same general rules reported for the case of the linear superposition state, even if the time-dependent fidelity has a fast-oscillating behavior. In the SC regime, the cavity-qubit interaction limits the maximum value that the fidelity can reach. Figure~\ref{coherSC}a shows the time evolution of the state transfer fidelity in the case in which $J=0.01$ and $g=0.01$, where complete transfer is not allowed. In contrast, in the USC regime, high values of the fidelity can be reached also when $g$ is larger than $J$. Figure~\ref{coherSC}b shows the coherent state transfer dynamics, for a case in which both the cavity-cavity and the cavity-qubit interactions are in the USC regime. The plot shows the fidelity $F$ over evolution time for a coherent state of amplitude $|\alpha|=1$. In this case, the fidelity can reach $F=0.9$, also if $g=2J$.

Notice that the considered cases, linear superpositions and coherent states, show the main features of the state transfer when considering the physics beyond the RWA.


\begin{thebibliography}{99}

\bibitem{CohenBook} 
C.~Cohen-Tannoudji, J.~Dupont-Roc, and G.~Grynberg, {\it Atom-Photon Interactions: Basic Processes and Applications} (WILEY-VCH Verlag GmbH \& Co. KGaA, Weinbeim, 2004).

\bibitem{QuDevices} 
T.~D.~Ladd, F.~Jelezko, R.~Laflamme, Y.~Nakamura, C.~Monroe, and J.~L.~O'Brien, Nature (London) {\bf 464}, 45 (2010).

\bibitem{CavityQED} 
S.~Haroche and J.-M.~Raymond, {\em Exploring the Quantum}  (Oxford Univ. Press Inc., New York, 2006); H.~Walther, B.~T.~H.~Varcoe, B.~G.~Englert, and T.~Becka, Rep. Prog. Phys. {\bf 69}, 1325 (2006).

\bibitem{OpLattices} 
I.~Bloch, Nature Phys. {\bf 1}, 23 (2005).

\bibitem{IonsReview}
R. Blatt and C. F. Roos, Nat. Phys. {\bf 8}, 277 (2012).

\bibitem{Plenio2006} 
M.~J.~Hartmann, F.~G.~S.~L.~Brand\~ao, and M.~B.~Plenio, Nature Phys. {\bf 2}, 849 (2006).

\bibitem{Hollenberg2006} 
A.~D.~Greentree, C.~Tahan, J.~H.~Cole, and L.~C.~L.~Hollenberg, Nature Phys. {\bf 2}, 856 (2006).

\bibitem{Bose2007} 
D.~G.~Angelakis, M.~F.~Santos, and S.~Bose, Phys. Rev. A {\bf 76}, 031805(R) (2007).

\bibitem{Rempe2012} 
S.~Ritter, C.~N\"{o}lleke, C.~Hahn, A.~Reiserer, A.~Neuzner, M.~Uphoff, M.~M\"{u}cke, E.~Figueroa,
J.~Bochmann, and G.~Rempe, Nature (London) {\bf 484}, 195 (2012).

\bibitem{Longo2010}
P.~Longo, P.~Schmitteckert, and K.~Busch, J. Opt. A: Pure Appl. Opt. {\bf 11}, 114009 (2009);
Phys. Rev. Lett. {\bf 104}, 023602 (2010);  Phys. Rev. A {\bf 83}, 063828 (2011).
  
\bibitem{Nori2008} 
L.~Zhou, Z.~R.~Gong, Y.-X.~Liu, C.~P.~Sun, and F.~Nori, Phys. Rev. Lett. {\bf 101}, 100501 (2008).

\bibitem{Zhang2012}
Z.~H.~Wang, Y.~Li, D.~L.~Zhou, C.~P.~Sun, and P.~Zhang, Phys. Rev. A {\bf 86}, 023824 (2012).

\bibitem{Blais04} 
A.~Blais, R.~S.~Huang, A.~Wallraff, S.~M.~Girvin, R.~J.~Schoelkopf, Phys. Rev. A {\bf 69}, 062320 (2004).

\bibitem{Chiorescu04} 
I.~Chiorescu, P.~Bertet, K.~Semba, Y.~Nakamura, C.~J.~P.~M.~Harmans, and J.~E.~Mooij, Nature (London) {\bf 431}, 159 (2004).

\bibitem{Wallraff04} 
A.~Wallraff, D.~I.~Schuster, A.~Blais, L.~Frunzio, R.-S.~Huang, J.~Majer, S.~Kumar, S.~M.~Girvin, and R.~J.~Schoelkopf, Nature (London) {\bf 431}, 162 (2004).

\bibitem{Houck2012a} 
A.~A.~Houck, H.~T\"{u}reci, and J.~Koch, Nature Phys. {\bf 8}, 292 (2012).

\bibitem{Hartmann2012} 
M.~Leib, F.~Deppe, A.~Marx, R.~Gross, and M.~Hartmann, New J. Phys. {\bf 14}, 075024 (2012).

\bibitem{Houck2012b} 
D.~Underwood, W.~E.~Shanks, J.~Koch, and A.~A.~Houck, Phys. Rev. A {\bf 86}, 023837 (2012). 

\bibitem{Bourassa09} 
J.~Bourassa, J.~M.~Gambetta, A.~A.~Abdumalikov, Jr., O.~Astafiev, Y.~Nakamura, and A.~Blais,
Phys. Rev. A {\bf 80}, 032109 (2009).

\bibitem{Niemczyk10} 
T.~Niemczyk, F.~Deppe, H.~Huebl, E.~P.~Menzel, F.~Hocke, M.~J.~Schwarz, J.~J.~Garc\'ia-Ripoll, D.~Zueco, T.~H\"ummer, E.~Solano, A.~Marx, and R.~Gross,  Nature Phys. {\bf 6}, 772 (2010).

\bibitem{Pol10} 
P.~Forn-D\'iaz, J.~Lisenfeld, D.~Marcos, J.~J.~Garc\'ia-Ripoll, E.~Solano, C.~J.~P.~M.~Harmans, and J.~E.~Mooij, Phys. Rev. Lett. {\bf 105}, 237001 (2010).

\bibitem{Braak2011} 
D.~Braak, Phys. Rev. Lett. {\bf 107}, 100401 (2011).

\bibitem{Tureci2012} 
M.~Schir\'{o}, M.~Bordyuh, B.~\"{O}ztop, and H.~E.~T\"{u}reci, Phys. Rev. Lett. {\bf 109}, 053601 (2012).

\bibitem{Saro2009} 
D.~Gerace, H.~E.~T\"{u}reci, A.~Imamoglu, V.~Giovannetti, and R.~Fazio, Nature Phys. {\bf 5}, 281 (2009).


\bibitem{Blatter2010}
 S.~Schmidt, D.~Gerace, A.~A.~Houck, G.~Blatter, and H.~E.~T\"ureci, Phys. Rev. B {\bf 82}, 100507(R) (2010).

\bibitem{Cleland2011} 
M.~Mariantoni, H.~Wang, R.~C.~Bialczak, M.~Lenander, E.~Lucero, M.~Neeley, A.~D.~O'Connell, D.~Sank, M.~Weides, J.~Wenner, T.~Yamamoto, Y.~Yin, J.~Zhao, J.~M.~Martinis, and A.~N.~Cleland, 
Nature Phys. {\bf 7}, 287 (2011).

\bibitem{footnote0}
The counter-rotating terms in the cavity-cavity coupling also induce photon generation from the vacuum. However, for the cases we consider ($J_{\ell} \lesssim g$) most of the non-conserving excitation effects come from the cavity-qubit USC regime.

\bibitem{SuzukiTrotter} 
  See, e.g., A.~T.~Sornborger and E.~D.~Stewart,  Phys. Rev. A {\bf 60}, 1956 (1999).

\bibitem{footnote}
In order to keep the Hilbert space dimension manageable in our simulations, we cut the maximum number of allowed photons per cavity to a finite amount $N_{\rm max} = 18$. We checked that such cutoff value induces an error that is negligible on the scale of our results, for $g / \omega\lesssim~1$. All Trotter time evolutions have been performed up to a time $t = 5000$.

\bibitem{BoxCount}
G.~De~Chiara, D.~Rossini, S.~Montangero, and R.~Fazio, Phys. Rev. A {\bf 72}, 12323 (2005).

\bibitem{footnote2}
For any curve, there exists a region of box lengths $\tau_{\rm min} < \tau < \tau_{\rm max}$ where $M \propto \tau^{\cal D}$. Outside this region, one either finds ${\cal D} = 1$ or ${\cal D} = 2$. The first equality holds for $\tau < \tau_{\rm min}$, and it is due to the coarse grain artificially introduced by numerical simulations. The second one is obtained for $\tau > \tau_{\rm max}$ and it is due to the finite length of the analyzed time series. The boundaries $\tau_{\rm min}$, $\tau_{\rm max}$ have to be chosen properly for any time series.

\bibitem{WolfVallone2012} 
 F.~A.~Wolf, F.~Vallone, G.~Romero, M.~Kollar, E.~Solano, and D.~Braak, Phys. Rev. A {\bf 87}, 023835 (2013).  

\bibitem{Solano2003} 
 E.~Solano, G.~S.~Agarwal, and H.~Walther, Phys. Rev. Lett. {\bf 90}, 027903 (2003).
  
\bibitem{Ballester2012} D. Ballester, G. Romero, J. J. Garc\'ia-Ripoll, F. Deppe, and E. Solano, Phys. Rev. X {\bf 2}, 023835 (2013).  

\end{thebibliography}
\end{document}